\documentclass[useAMS,usegraphicx]{mn2e}
\usepackage{rotate}
\usepackage{times}
\newif\ifAMStwofonts
\AMStwofontstrue

\def\gs{\mathrel{\hbox{\rlap{\hbox{\lower4pt\hbox{$\sim$}}}\hbox{$>$}}}}
\def\ls{\mathrel{\hbox{\rlap{\hbox{\lower4pt\hbox{$\sim$}}}\hbox{$<$}}}}

\def\einstein{{\it Einstein}}

\def\asca{{\it ASCA}}
\def\sax{{\it BeppoSAX}}

\def\xmm{{\it XMM-Newton}}
\def\ginga{{\it Ginga}}

\def\et{{et al.\ }}
\def\mcg{{MCG--6-30-15}}
\def\ngc4051{{NGC~4051}}
\def\pg1535{{PG~1535+547}}
\def\iras13224{{IRAS~13224--3809}}
\def\1h{{1H~0707--495}}
\def\phl{{PHL~1092}}
\def\3c{{3C~273}}

\def\oiii{{[\rm O~\textsc{iii}]}}

\def\feii{{\rm Fe~\textsc{ii}}}
\def\hbeta{{$\rm H\beta$}}

\def\A{{\rm\thinspace \AA}}
\def\cm{{\rm\thinspace cm}}
\def\erg{{\rm\thinspace erg}}
\def\eV{{\rm\thinspace eV}}

\def\Hz{{\rm\thinspace Hz}}

\def\keV{{\rm\thinspace keV}}
\def\km{{\rm\thinspace km}}

\def\s{{\rm\thinspace s}}

\def\ps{{\rm\thinspace s^{-1}}}

\def\ergpscmps{\hbox{$\erg\cm^{-2}\s^{-1}\,$}}

\def\ergpscmpspHz{\hbox{$\erg\cm^{-2}\s^{-1}\Hz^{-1}\,$}}

\def\kmps{\hbox{$\km\ps\,$}}

\def\pscm{\hbox{$\cm^{-2}\,$}}

\title[Investigation of complex NLS1]
      {
Investigating the nature of narrow-line Seyfert 1 galaxies with high-energy
spectral complexity
      }

\author[L. C. Gallo]
       {L. C. Gallo  \\
Max-Planck-Institut f\"ur extraterrestrische Physik, Postfach 1312, 85741 Garching, Germany \\
Institute of Space and Astronautical Science, Japan Aerospace Exploration Agency, Yoshinodai 3-1-1, Sagamihara, Kanagawa 229-8510, Japan \\
}
\date{Accepted. Received. }
\pagerange{\pageref{firstpage}--\pageref{lastpage}}
\pubyear{2005}
\begin{document}
\maketitle
\label{firstpage}

\begin{abstract}
With the commissioning of \xmm\ came the discovery of $2.5-10\keV$ spectral 
complexity in some narrow-line Seyfert 1 galaxies (NLS1).  This
high-energy complexity can be manifested as sharp, spectral drops or gradual
curvature in the spectrum.  Models which are normally considered are ionised 
reflection 
and partial-covering.  In this work, we define two samples of NLS1: a complex
sample whose members exhibit high-energy complexity (C-sample), and a general sample 
of NLS1 whose $2.5-10\keV$ spectra do not strongly deviate from a simple power law  
(S-sample).  We than compare
multiwavelength parameters of these two samples to determine if there are
any distinguishing characteristics in the complex NLS1.  
Considering historical light curves of each object we find that the C-sample
is representative of NLS1 in a low X-ray flux state, whereas the members
of the S-sample appear to be in a typical flux state.
Moreover, from measurements
of $\alpha_{ox}$ with contemporaneous UV/X-ray data, we find that the 
C-sample of NLS1 
appear X-ray weaker at the time of the observation.
For two NLS1 in the C-sample multi-epoch 
measurements of
$\alpha_{ox}$ are available and suggest that $\alpha_{ox}$ approaches more 
normal values as the complexity between $2.5-10\keV$ diminishes.  
This implies that a source could transit from one sample to the other
as its X-ray flux varies.
Secondly, there are indications that the C-sample sources, on average,  
exhibit stronger optical \feii\ emission, with the three most extreme
(\feii/\hbeta$>1.8$) \feii\ emitters all displaying
complexity in the $2.5-10\keV$ band.  It is an intriguing possibility that we may be
able to identify X-ray complex NLS1 based on the extreme strength of the more
easily observable optical \feii\ emission.  
However, it is not clear if the possible connection between \feii\ strength 
and spectral complexity is due to the \feii\ producing mechanism or because
strong \feii\ emitters may exhibit the greatest variability and consequently
more likely to be caught in an extreme (low) flux state.
Based on the current analysis it we can not straightforwardly dismiss
absorption or reflection as the cause of the X-ray complexity;
by considering the multiple UV/X-ray observations of \1h\ (a C-sample member),
we discuss a possible method of distinguishing the two models provided further
UV/X-ray observations.

\end{abstract}



\section{Introduction}
\label{sect:intro}

The importance of narrow-line Seyfert 1 galaxies (NLS1) became apparent on 
the discovery of the `primary eigenvector' (PC1; Eigenvector 1) in
the Boroson \& Green (1992) principle component analysis of PG quasars.
PC1 showed a strong anticorrelation between the strengths of
\oiii\ and \feii\ in the optical spectra, with NLS1 as a class showing the
strongest \feii\ emission and weakest \oiii.
While the physical driver of PC1 is still debated, there are strong
indications that the fraction of the Eddington luminosity ($L/L_{Edd}$)
at which the object is emitting at 
is responsible, implying that NLS1 are relatively
high accretion rate systems.
Since there are no significant differences between NLS1 and broad-line Seyfert
1 galaxies (BLS1) in their X-ray, optical, or bolometric luminosities 
(Grupe \et 2004), it naturally follows from the condition of higher accretion
rates that NLS1 possess a lower mass black hole compared to BLS1 of similar 
luminosity.
Black hole mass measurements confirm 
lower mass black holes in NLS1 (Wandel,
Peterson \& Malkan 1999; Peterson \et 2000; Grupe \& Mathur 2004).

The strong \feii\ emission in NLS1 is prevalent in the UV (e.g.
Laor \et 1997; Constantin \& Shields 2003), optical
(e.g. Boroson \& Green 1992),
and infrared (e.g. Rodr\a'\i guez-Ardila \et 2002).  The observed line widths
and absence of forbidden emission suggests that \feii\ is formed
in the dense BLR, but photoionisation models cannot account for all of the 
\feii\ emission.
The `\feii\ discrepancy' remains unsolved, though models which
consider non-radiative heating (probably due to shocks produced in outflows),
with an overabundance of iron are promising 
(see Collin \& Joly 2000 for a review).

Perhaps the most interesting characteristics of NLS1 are manifested in the
X-ray regime.  The origin of the strong soft excess emission below
about $1\keV$, known since \einstein\ observations (e.g. Puchnarewicz \et 1992;
see also Boller, Brandt \& Fink 1996), is still debated
(e.g. Mineshige \et 2000; Gierlinski \& Done 2004; Crummy \et 2005); 
as is the nature of the extreme, rapid variability.
With \asca, came the discovery that the $2-10\keV$ spectrum in NLS1 also 
appeared steeper than in BLS1 (Brandt, Mathur \& Elvis 1997).
The result could be understood as arising from significant Compton cooling of 
the accretion disc corona due to the strong soft X-ray excess found in most 
NLS1.  The lower energy gain per scattering (smaller Compton $y$ parameter)
would yield steeper, hard X-ray slopes.

With the high-sensitivity of \xmm\ came what may be the most interesting
discovery with regards to NLS1, and that was the presence of a sharp, spectral
drop at $E=7\keV$ in \1h\ (Boller \et 2002).  Since then, a number of drops at
$E\gs7\keV$ (or more generally `high-energy curvature') have been observed in 
several NLS1 (e.g. Pounds \et 2003, 2004; Longinotti \et 2003;
Boller \et 2003; Uttley \et 2004; Reeves \et 2004).
The exact nature of this behaviour, which seems to be a characteristic of NLS1,
is uncertain.  Two models which appear probable are partial-covering
(e.g. Tanaka \et 2004 and references within) and reflection 
(e.g. Fabian \et 2002).  

In terms of partial-covering, the $\sim7\keV$ drop is produced by absorption 
of the continuum emission by a dense material, which only partly obscures the
primary emitter.  This can, in principle, explain the absence of other 
absorption features (e.g. intrinsic cold absorption, Fe~L absorption, 
fluorescence emission), and if
the absorber is allowed to be in radial motion, it can possibly account for
the various edge energies which are seen (e.g. Gallo \et 2004a).
It is important to realise that partial-covering does not describe the nature
of the primary continuum source.  Although a simple blackbody plus power law
continuum is often assumed, other physical descriptions of the primary 
continuum are not dismissed, nor do they discriminate against
partial-covering.

Reflection of the power law continuum source off the cold accretion
disc can also adequately describe the X-ray spectra of NLS1 
(e.g. Fabian \et 2004).  In this case, the sharp drop at high energies
is the blue wing of a
relativistically broadened iron line.  In combination with light bending
effects close to the black hole (e.g. Miniutti \& Fabian 2004),  the 
reflection model nicely describes the shape of the X-ray continuum and the
principle of `reflection dominated' spectra.

It stands to reason that regardless of the correct model, the process may
be ubiquitous in NLS1 and probably in active galactic nuclei (AGN).  
By varying the prevalence of the
physical process (e.g. the degree of absorption or amount
of reflection) one can potentially describe the different types of 
X-ray spectra that are observed.  
Objects, such as \iras13224\ or \1h, manifest `the process' significantly; thus
exhibiting sharp and deep spectral drops.  Other NLS1, like NGC~4051, demonstrate
the process only moderately, only displaying gentle curvature over the hard
X-ray band.  In most other NLS1, the process is minimal and likely not detected.

In this work we examine what, if anything, is unique about the NLS1
which appear to possess high-energy complexity.  
Leighly \& Moore (2004) began to address this issue when they 
examined how the UV properties of \iras13224\ and \1h\ were remarkable compared
to other NLS1.  They found that \iras13224\ and \1h\ possessed weaker emission
lines, significant asymmetry in the C~\textsc{iv} profile and steeper spectra.
They concluded that many of the differences in the two NLS1
could be explained in terms of radiative line driven wind models
(e.g. Proga, Stone \& Kallman 2000; Leighly 2004).
Here, we continue to probe this issue by investigating multiwavelength
properties of a sample of NLS1, which includes a larger number of `extreme'
objects.

The remainder of the paper is organised as follows.  In the next section
we define a complex sample of NLS1, which appear 
to exhibit high-energy complexity in their \xmm\ (Jansen \et 2001) spectra;
and a general sample of NLS1, which do not show significant evidence of
the complexity described here.  
In addition, the X-ray data processing is described.  
The dependence of X-ray flux state in defining the samples is
investigated in Section~3.
In Section~4, the calculation of the optical-to-X-ray spectral index from 
contemporaneous UV and X-ray data are described and presented.  In Section~5
the multiwavelength parameters obtained from the literature (or estimated
by us) are compared and our findings are highlighted.  We discuss our results in
Section~\ref{sect:disc} and give our summary in Section~\ref{sect:conc}.
A value for the Hubble constant of $H_0$=$\rm 70\ km\ s^{-1}\ Mpc^{-1}$ and
a standard flat cosmology with $\Omega_{M}$ = 0.3 and $\Omega_\Lambda$ = 0.7
were adopted throughout.

\section{X-ray data processing and sample definition}
\label{sect:samp}

\begin{table*}
\begin{center}
\caption{The X-ray data. 
The complete sample is divided into two sub-samples as described in the text.
The information given is the:
object name (column 1);
observation data (column 2) and \xmm\ revolution (column 3);
total good exposure for the pn (column 4);
observed $2.5-10\keV$ flux ($\times10^{-11}\ergpscmps$) corrected for Galactic absorption (column 5);
intrinsic $2\keV$ flux density ($\times10^{-30}\ergpscmpspHz$) (column 6);
null hypothesis from fitting the base model to
the $2.5-10\keV$ data (column 7);
}
\begin{tabular}{ccccccc}
\multicolumn{7}{c}
   {General NLS1 Sample (S-sample)} \\
\hline
(1) & (2) & (3) & (4) & (5) & (6) & (7)  \\
Source & Date       & \xmm\ & pn exposure & $2.5-10\keV$  & $2\keV$      & Null       \\ 
       & year.mm.dd &  rev. & (ks)         & flux       & flux density & Hypothesis \\
\hline
Mrk~766$^1$       & 2000.05.20 & 0082 & 26.0 & 1.33 & 19.08 & 0.757 \\
Mrk~359           & 2000.07.09 & 0107 &  6.9 & 0.51 &  6.69 & 0.418 \\ 
Mrk~1044          & 2002.07.23 & 0480 &  5.5 & 0.57 &  9.23 & 0.578 \\ 
Akn~564$^{1,2}$   & 2000.06.17 & 0096 &  7.4 & 1.86 & 46.39 & 0.992 \\ 
Mrk~896           & 2001.11.15 & 0355 &  7.2 & 0.31 &  5.08 & 0.339 \\ 
Mrk~335           & 2000.12.25 & 0192 & 28.5 & 1.20 & 22.53 & 0.731 \\ 
NGC~7158$^2$      & 2001.11.27 & 0361 &  4.5 & 0.02 &  0.19 & 0.890 \\ 
Mrk~493           & 2003.01.16 & 0568 & 13.3 & 0.31 &  5.95 & 0.629 \\ 
I~Zw~1            & 2002.06.22 & 0464 & 18.6 & 0.68 & 14.12 & 0.182 \\ 
Ton~S180$^1$      & 2000.12.14 & 0186 & 20.6 & 0.39 &  9.00 & 0.292 \\ 
PG~1448+273       & 2003.02.08 & 0580 & 18.3 & 0.16 &  3.71 & 0.697 \\ 
UGC~11763         & 2003.05.16 & 0629 & 26.2 & 0.31 &  3.46 & 0.102 \\ 
RX~J0323.2--4931  & 2003.08.16 & 0675 & 25.1 & 0.11 &  1.90 & 0.103 \\ 
Mrk~478$^1$       & 2003.01.07 & 0564 & 18.2 & 0.15 &  2.98 & 0.972 \\ 
II~Zw~177$^2$     & 2001.06.07 & 0274 &  8.7 & 0.08 &  2.09 & 0.604 \\ 
IRAS~13349+2438   & 2000.06.20 & 0097 & 30.4 & 0.19 &  3.08 & 0.162 \\ 
PKS~0558--504$^1$ & 2000.10.10 & 0153 &  8.0 & 1.03 & 20.75 & 0.519 \\ 
PG~1115+407$^2$   & 2002.05.17 & 0446 & 15.0 & 0.10 &  2.35 & 0.746 \\ 
Nab~0205+024$^2$  & 2002.07.23 & 0480 & 21.9 & 0.13 &  3.18 & 0.320 \\ 
PG~2233+134$^2$   & 2003.05.28 & 0635 &  7.5 & 0.05 &  1.15 & 0.713  \\ 
E~1346+266$^2$    & 2003.01.13 & 0567 & 49.2 & 0.005&  0.47 & 0.464 \\ 

\hline
\multicolumn{7}{c}
   {Complex NLS1 Sample (C-sample)} \\
\hline
NGC~4051$^1$      & 2002.11.22 & 0541 & 44.7 & 0.59 &  4.98 & $<0.01$ \\ 
PG~1535+547$^2$   & 2002.11.03 & 0531 & 20.1 & 0.12 &  0.12 & $<0.01$ \\ 
1H~0707--495$^1$  & 2000.10.21 & 0159 & 38.1 & 0.04 &  0.39 & $<0.01$ \\ 
IRAS~13224--3809  & 2002.01.19 & 0387 & 54.3 & 0.04 &  0.84 & $<0.01$ \\ 
PG~1211+143       & 2001.06.15 & 0278 & 49.5 & 0.27 &  3.24 & $<0.01$ \\ 
PG~1402+261       & 2002.01.27 & 0391 &  9.1 & 0.15 &  3.30 & $0.09$  \\ 
PHL~1092$^2$      & 2003.01.18 & 0570 & 19.8 & 0.02 &  0.31 & $0.20$   \\

\hline
\label{tab:xmm}
\end{tabular}

\medskip
\raggedright
FOOTNOTES:
(1) Multiple \xmm\ observations exist.
(2) Adding an unresolved Gaussian profile was not an improvement over a simple power law fit.
\end{center}
\end{table*}

\begin{figure}
\rotatebox{270}
{\scalebox{0.32}{\includegraphics{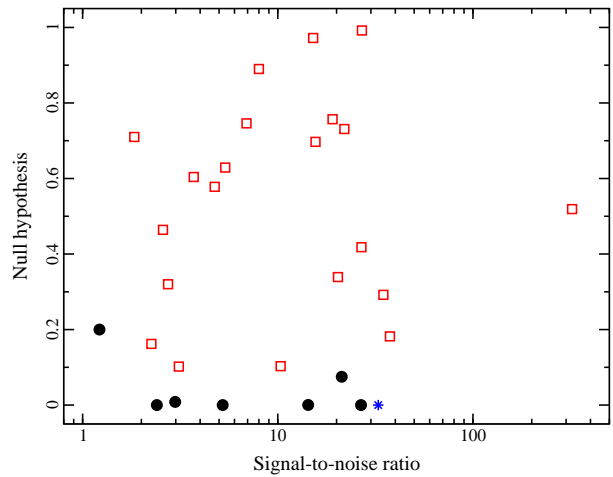}}}
\caption{The null hypothesis of a power law plus Gaussian profile model to the
intrinsic $2.5-10\keV$ spectrum of each NLS1 in the sample plotted against
$5-10\keV$ SN.  The filled (black) circles mark the NLS1 identified as
complex (C-sample; see text for details).  The open (red) squares are the general NLS1
sample (S-sample).  \mcg\ is shown as a blue star for illustrative purposes.
}
\label{fig:samp}
\end{figure}
In defining a sample we considered all known NLS1 that have been observed
on-axis and in imaging mode with the EPIC pn camera (Str\"{u}der \et 2001) on
\xmm.  The pn was selected to make use of what is most likely the highest
signal-to-noise (SN), $2.5-10\keV$ rest-frame spectrum of these objects,
currently available.  At the time we commenced this study, there were more
than 50 observations of 37 NLS1-type objects (including quasars and
low-luminosity sources).

For each observation, the Observation Data Files (ODFs) were processed to
produce calibrated event lists using the \xmm\ Science Analysis System
({\tt XMM-SAS v6.1.0}) and the most recent calibration files.
Unwanted hot, dead, or flickering pixels were removed as were events due to
electronic noise.  Event energies were corrected for charge-transfer
losses, and time-dependent EPIC response matrices were generated using the
{\tt SAS} tasks {\tt ARFGEN} and {\tt RMFGEN}.
Light curves were extracted from these event lists to search for periods of
high background flaring, which were then removed.
The source plus background photons were extracted from a source-centred
circular region with a radius of 35$^{\prime\prime}$.  The background was
selected from an off-source region with a larger radius and appropriately
scaled to the source region.
Single and double events were selected for the pn spectra.
Pile-up was examined in all spectra and when corrections were necessary
the inner pixels of the source extraction region were ignored.

On processing of the spectral data, the sample was further refined by excluding
observations in which: (i) the SN in the intrinsic $5-10\keV$ range was
$\ls 1.2$, or (ii) the intrinsic $5-9\keV$ spectrum was not source dominated.
Finally, if more than one observation of the same object was available, the
highest SN data were used.  This filtering process resulted in the sample
of 28 NLS1, which possess high-quality, intrinsic $2-10\keV$ spectra
(Table~\ref{tab:xmm}).

The purpose of this study is to focus on the nature of NLS1 which exhibit
complexity in their intrinsic $\sim 2.5-10\keV$ spectra.  To identify
these objects, each spectra above $2.5\keV$ (rest-frame) was fitted with a
baseline model composing of a power law modified by Galactic absorption
(Dickey \& Lockman 1990) plus, if necessary, an
unresolved ($\sigma\ls150\eV$) Gaussian profile between $6.4-7\keV$.
The intention of this baseline model was not to determine the `best-fit' to
the spectrum, but to identify the objects which, to first order, could
not be explained by the most basic AGN model.

Sources for which the null hypothesis of this baseline model was
$<0.10$ were marked as NLS1 with high-energy complexity.
We purposely adopted a low value for the null hypothesis to identify the
most extreme cases which, in turn, should make it easier to identify other
extreme characteristic of the sample.  Six NLS1 (PG~1211+143, \iras13224,
\1h, \pg1535, PG~1402+261 and \ngc4051) were categorised as having high-energy
complexity. In addition, \phl\ was also included in this sub-sample
because it exhibited a very flat, and likely unphysical, $2.5-10\keV$ power
law ($\Gamma\approx1.55$), even though the baseline model null hypothesis was
0.20.  In total, seven objects make up the `complex NLS1 sample' (hereafter C-sample).  
The remaining twenty-one sources compose the `general (simple) NLS1 sample' 
(hereafter S-sample) (see Table~\ref{tab:xmm}).

Notably absent from the C-sample is the borderline NLS1, IRAS~13349+2438
(FWHM(\hbeta)$\approx2800\kmps$, but strong \feii\ emission; Grupe \et 2004),
which clearly shows high-energy complexity (Longinotti \et 2003), but does not
satisfy the rather strict criteria established here.

In Figure~\ref{fig:samp} the null hypothesis is plotted against the
$5-10\keV$ SN for the objects in the sample.  As will be adopted throughout,
the filled circles will mark the members of the C-sample and the open squares will
identify the S-sample population.  
The distribution of SN is comparable for the C- and S-sample.  Therefore,
even though identifying X-ray
spectral complexity obviously depends on SN, it does not appear that it
will be a significant bias in selecting our sample.
For illustrative purposes the pn data for
\mcg\ during revolution 302 are also included (blue star).
We chose to include \mcg\ as it is perhaps the most obvious
display of high-energy spectral complexity in a type 1 AGN.

\section{X-ray flux dependence on sample definition }
\label{sect:om}

We considered what effect X-ray variability could have on
the sample definition by examining the historic $2-10\keV$
light curve for each object.
Specifically, we compared the \xmm\ X-ray fluxes measured for each 
object with past measurements from other missions.  

\asca\ fluxes were
available either from Vaughan \et (1999) or the TARTARUS archive
for ten of the objects in the S-sample.  
Five objects also had \sax\ measurements (Comastri 2000) resulting in
an X-ray flux measurement for eleven NLS1 in the S-sample (four having
\sax\ and \asca\ measurements).
We found that for ten of them
the flux varied by less than 50 per cent, although more typically on the $10-15$ per cent
level.  This degree of variability is not abnormal for NLS1 on hourly
time scales; therefore it seems that most members of the S-sample
are in a typical flux state (of course, this is based on limited observations).
The \xmm\ flux of IRAS~13349+2438 diminished by more than $\sim70$ per cent since
the two \asca\ observations.  During the \sax\ observation, IRAS~13349+2438
was another $\sim70$ per cent dimmer then during the \xmm\ pointing.
Due to our strict criteria, IRAS~13349+2438 has not been included in our C-sample,
but it does appear to possess some high-energy complexity 
(Longinotti \et 2003).  It could be that in even lower flux states, such
as that observed with \sax, the spectrum would appear even more
complicated.

For three members of the C-sample, \1h, NGC~4051 and PG~1211+143, it 
can be established from long-term, multi-mission light curves
that the objects were in low-flux states
during the \xmm\ observations (see Leighly \et (2002) for \1h,
Uttley \et (2003, 2004) for NGC~4051, and Yaqoob \et (1994) for 
PG~1211+143\footnote{The X-ray flux of PG~1211+143 is also low 
compared to the \ginga\ observations (Lawson \& Turner 1997). 
}).  
Similarly, from a 10-day long
\asca\ observation (Dewangan \et 2000) and comparison with the \sax\ observation
(Comastri 2000) it can be shown that IRAS~13224--3809 was also in an X-ray 
low-flux state during the \xmm\ observation.

Data are sparse for PG~1535+547, PG~1402+261 and PHL~1092.  
Compared only to the \asca\ fluxes, the \xmm\ flux for PG~1535+547 was comparable
while for PG~1402+261 it was $\sim40$ per cent lower.
For PHL~1092
the \xmm\ flux was about $50$ per cent lower than during the \asca\
observation.  It is difficult to determine if this indicates a low-flux
state for PHL~1092.  Variations of this order are typical on hourly
time scales in the $0.1-2\keV$ range (Brandt \et 1999; Gallo \et 2004b);
however the $2-10\keV$ variability is found to be negligible over
similar time scales (Gallo \et 2004b).

If not convincing, it appears rather suggestive that the S- and
C-samples defined in this work
are, by coincidence, representative of NLS1 in X-ray typical- and low-flux 
states, respectively.
Therefore, it also follows that objects could transit from one sample
to another depending on their X-ray flux state.

\section{Optical and UV properties }
\label{sect:om}

An advantage afforded with \xmm\ is the availability of optical and
UV data obtained with the Optical Monitor (OM; Mason \et 2001) simultaneously
with X-ray observations.
For many of the NLS1 observations, OM imaging data were also collected at UV
or optical wavelengths.  Therefore, whenever possible, we determined
rest-frame $2500\A$ luminosities and optical-to-X-ray spectral indices
($\alpha_{ox}$).

The standard definition of $\alpha_{ox}$ is:
$
\alpha_{ox} = log(f_{x}/f_{uv})/log(\nu_{x}/\nu_{uv}),
$
where $f_{x}$ and $f_{uv}$ are the intrinsic flux densities at $2\keV$ and 
$2500\A$,
respectively; and $\nu_x$ and $\nu_{uv}$ are the corresponding frequencies.
In many of the observations the rest-frame $2500\A$ was directly
observed within one of the OM broad-band filters.  Consequently, the flux density
at $2500\A$ was estimated from the source count rate in that filter (Chen 2004).
However, when this was not the case it was necessary to extrapolate from some
measured UV wavelength ($u$) to $2500\A$.  The modified expression for 
$\alpha_{ox}$ is then:
$
\alpha_{ox} =log(f_{x}/f_{u})/log(\nu_{x}/\nu_{u}) +
log(f_{u}/f_{uv})/log(\nu_{u}/\nu_{uv}),
$
where the second term on the right-hand side is the spectral slope
between the measured UV flux density at $u$ and $2500\A$.
If measured by Constantin \& Shields (2003), the source-specific value for
the UV spectral slope was used (the value shown in Table~\ref{tab:samp}).
If the source-specific value was not known we adopted the spectral slope of
the composite SDSS quasar spectrum between $1300-5000\A$ ($\alpha_u = -0.44$;
Vanden Berk \et 2001).

The $2500\A$ flux density was measured from the available pre-processed
pipeline products for each source.  For most observations, multiple images were
taken in the same filter, in which case the average source count rate was used
to estimate the flux densities (Chen 2004).  For Mrk~1044, Mrk~896 and Mrk~493,
images were only made in the $V$-band ($5100-5800\A$).  For these observations
an extrapolation to $2500\A$ was not done given that the optical spectra of
AGN show a break in the continuum slope at about these wavelengths, which would
introduce systematic uncertainties in the measurements of these three NLS1.
Host-galaxy contribution was not excluded; thus will introduce a level
of uncertainty in the reported values.

\begin{table*}
\begin{center}
\caption{Optical and UV properties.
The source name is given in column 1.
The other parameters are the: $2500\A$ flux density 
($\times10^{-26}\ergpscmpspHz$) 
(column 2);
\feii/\hbeta (column 3); FWHM of \hbeta\ ($\kmps$) (column 4);
UV flux density ($\times10^{-26}\ergpscmpspHz$) 
(column 5);
$\sim1100-4000\A$ spectral index ($F \propto \nu^{\alpha_u}$) (column 6);
and $\alpha_{ox}$ (column 7).  Values in columns 5 and 6 are listed
only when the $2500\A$ flux density (column 2) was not directly
available from the data.
}
\scalebox{1.0}{
\begin{tabular}{ccccccc}
\multicolumn{7}{c}
   {General NLS1 Sample (S-sample)} \\
\hline
(1)  & (2) & (3) & (4) & (5) & (6) & (7)  \\
Source & $f_{uv}$ & \feii/\hbeta & FWHM(\hbeta) & $f_u$ & $\alpha_u$ & $\alpha_{ox}$ \\
\hline
Mrk~766          & 0.57 & 1.56  & 1100  & --   & --      & $-0.958$ \\ 
Mrk~359          & 1.41 & 0.50  &  900  & 1.33 & $-0.44$ & $-1.276$ \\
Mrk~1044         & --   & 0.77  & 1310  & --   & --      & --       \\ 
Akn~564          & 1.31 & 0.67  &  865  & --   & --      & $-0.941$ \\
Mrk~896          & --   & 0.50  & 1135  & --   & --      & --       \\
Mrk~335          & 5.43 & 0.62  & 1710  & 5.27 & --0.64  & $-1.298$ \\
NGC~7158         & --   & --    & 2100  & --   & --      & --       \\
Mrk~493          & --   & 1.16  &  800  & --   & --      & --       \\
I~Zw~1           & 1.61 & 1.47  &  1240 & 1.20 &  --1.75 & $-1.170$ \\
Ton~S180         & 4.40 & 0.90  &  970  & 3.88 & --0.76  & $-1.417$ \\
PG~1448+273      & 0.98 & 0.94  & 1330  & --   & --      & $-1.407$ \\
UGC~11763        & 2.61 & 0.63  &  2210 & --   &  --     & $-1.489$ \\
RX~J0323.2--4931 & 0.17 & 0.65  &  1680 & --   &  --     & $-1.131$ \\
Mrk~478          & 2.35 & 0.97  & 1630  & --   & --      & $-1.495$ \\
II~Zw~177        & --   & --    & 1176  & --   & --      & --       \\
IRAS~13349+2438  & 1.06 & 1.25  &  2800 & 0.54 & --3.23  & $-1.359$ \\
PKS~0558--504    & --   & 1.60  & 1250  & --   & --      & --       \\
PG~1115+407      & 0.82 & 0.98  & 1740  & 0.84 & --0.44  & $-1.371$ \\
Nab~0205+024     & 1.02 & 0.62  & 1050  & 1.03 & --0.44  & $-1.340$ \\
PG~2233+134      & 0.53 & 0.89  & 1740  &  --  & --      & $-1.407$ \\
E~1346+266       & 0.03 & 0.98  & 1840  & --   & --      & $-1.048$ \\

\hline

\multicolumn{7}{c}
   {Complex NLS1 Sample (C-sample)} \\
\hline
NGC~4051         & 3.64 & 0.94  & 1170 & --    &  --     & $-1.484$ \\
PG~1535+547      & --   & 0.47  & 1480 & --    &  --     & --       \\
1H~0707--495     & 1.54 & 2.77  & 1000 & 1.44  &  --0.46 & $-1.763$ \\
IRAS~13224--3809 & 0.65 & 2.40  &  650 & 0.60  &  --0.44 & $-1.536$ \\
PG~1211+143      & 2.89 & 0.50  & 1900 & --    &  --     & $-1.517$ \\
PG~1402+261      & 1.87 & 1.10  & 1623 & --    &  --     & $-1.441$ \\
PHL~1092         & 0.36 & 1.81  & 1300 & 0.30  &  --0.44 & $-1.560$ \\
\hline

\label{tab:samp}
\end{tabular}
}

\end{center}
\end{table*}

In addition, FWHM(\hbeta) and \feii/\hbeta\ ratios were collected from the
literature for as many objects as possible.  These are also included in 
Table~\ref{tab:samp}.

\section{Multiwavelength parameters}
\label{sect:prop}

In order to probe the multiwavelength behaviour of these NLS1 we
collected various parameters from the literature such as:
radio loudness, optical emission line strengths, C~\textsc{iv} strength,
black hole mass and Eddington luminosity ratios.
Comparing the two samples in all parameter spaces
possible with these data reveled no clear difference between the C- and S-sample;
thus these parameters will not be considered further.

\subsection{Contemporaneous X-ray and UV properties}

\begin{figure}
\rotatebox{270}
{\scalebox{0.32}{\includegraphics{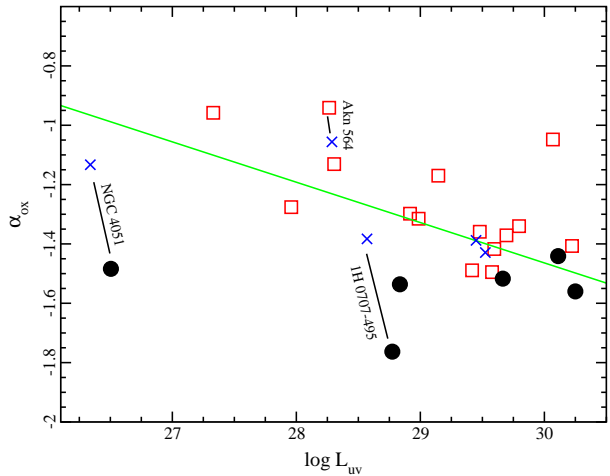}}}
\caption{$\alpha_{ox}$ as a function of $2500\A$ luminosity
measured from contemporaneous X-ray and UV data.
The filled (black) circles mark the C-sample NLS1 and the
open (red) squares are the S-sample NLS1.
For comparison, the relation $\alpha_{ox}=-0.136L_{uv}+2.616$, found for
radio-quiet type 1 AGN
(Strateva \et 2005) is shown as a green, solid line.
The general NLS1 sample appears to follow the Strateva relation well, but the
complex objects are X-ray weak.
Multiple \xmm\ observations exist for some of the
objects in our sample.  These are shown as blue crosses. The $\alpha_{ox}$
variations displayed by \1h\ and NGC~4051 indicate significant
changes in the X-ray flux and more moderate variations in the UV.
Akn~564 appears excessively X-ray strong, but
this is likely due to reddening of the UV spectrum (Crenshaw \et 2002).
}
\label{fig:aox}
\end{figure}

Measurements of $\alpha_{ox}$ from simultaneous UV and X-ray data clearly
demonstrate a difference in the slope of the optical-X-ray continuum in
the two samples (Figure~\ref{fig:aox}).  The average index for the 
S-sample is $-1.268\pm0.179$ compared to $-1.550\pm0.112$ for the 
C-sample.  
A Kolmogorov-Smirnov (KS) test comparing the two distributions
yields a probability of $<0.1$ per cent that they are drawn from 
the same sample.

In Figure~\ref{fig:aox} we have plotted $\alpha_{ox}$ as a function of
$2500\A$ monochromatic luminosity.  For comparison, the UV luminosity
dependence of $\alpha_{ox}$ for radio-quiet, type 1 AGN 
($\alpha_{ox}=-0.136L_{uv}+2.616$) 
is also shown (Strateva \et 2005). 
The S-sample appears to
follow this relationship relatively well, but most of the C-sample
fall below the Strateva relation.
In combination with the fact that the C-sample objects are in X-ray
low-flux states (Section~3) this 
clearly demonstrates X-ray weakness in these complex objects during
the observations.

For five NLS1 in the overall sample, at least two X-ray observations with
contemporaneous OM data are available.
Three of these objects (Ton~S180, Mrk~478, Akn~564) are from the 
S-sample, the other two (\1h, NGC~4051) are from the C-sample.
The epoch-specific $\alpha_{ox}$ and $L_{uv}$ were calculated for each of
these observations.  The results are plotted in Figure~\ref{fig:aox} as
crosses.  For the objects in the S-sample, the fluctuations in
$\alpha_{ox}$ are small and perhaps even negligible if a complete error
analysis was considered.

In contrast, the $\alpha_{ox}$ variations are much more significant for the
two objects in the C-sample (\1h\ and NGC~4051).  Figure~\ref{fig:aox}
also indicates that the variations in $\alpha_{ox}$ are driven by changes in
the X-ray flux, but that the UV flux also changes.
For example, during the second observation \1h\ was about six times brighter
in the X-rays, while its UV flux diminished by $\sim50$ per cent.
Increasing X-ray flux and decreasing UV flux was also seen in NGC~4051.

We also note that in the second observations of \1h\ and NGC~4051,
both objects portray simpler high-energy spectra.  That is, the
null hypothesis for a power law plus Gaussian fit for \1h\ and NGC~4051 at these
epochs was 0.176 and 0.456, respectively.  By our definition, neither object
would have been considered `complex' at those times,
suggesting that when measured in ``typical'' flux states, these source
also follow the Strateva relation.

\subsection{Optical properties}

\begin{figure}
\scalebox{0.32}{\includegraphics[angle=270]{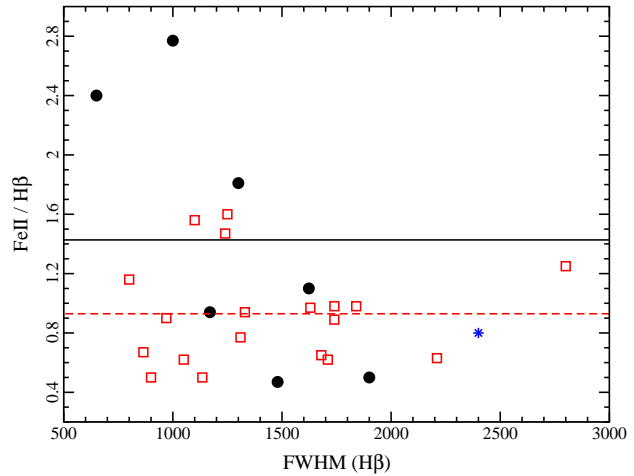}}
\caption
{\label{fig:feii}
The \feii/\hbeta\ ratio is plotted against the FWHM(\hbeta).
The symbols are as defined previously.  
The average
ratio for the S-sample (open squares) and the C-sample
(filled circles) are marked by a dashed and solid line, respectively.
Although there is considerably more scatter in the average ratio of the
complex NLS1 sample, the objects appear to possess much stronger \feii\
emission compared
to \hbeta.  The average ratios are $0.93\pm0.34$ (general sample) and
$1.43\pm0.91$ (complex sample).  The sample is limited, but it appears
the most extreme \feii\ emitters (e.g. \feii/\hbeta$\gs1.7$)
also display complexity in their high-energy spectra.
}
\end{figure}

In Figure~\ref{fig:feii}, the \feii/\hbeta\ ratio is plotted against the
FWHM(\hbeta).
The average ratio for the S-sample is $0.93\pm0.34$ with modest
scatter as indicated by the reported standard deviation.  The average of the
C-sample is $1.43\pm0.91$ with considerable more scatter.
The two samples clearly overlap.  
From a KS-test it is determined that the probability that they are drawn
from the same sample is 8 per cent.
Although not highly significant there are indications that the strongest \feii\
emitters are the ones that also display high-energy spectral complexity.

\section{Discussion}
\label{sect:disc}

\subsection{X-ray weak NLS1}
According to Figure~\ref{fig:aox}, objects that possess
complexity in their $2.5-10\keV$ spectrum appear to be X-ray weaker
(i.e. they are found below the average $L_{uv}-\alpha_{ox}$ relation).
At least in two of the objects (NGC~4051 and \1h) for which multiple,
simultaneous, X-ray/UV observations are available, extreme epoch-to-epoch
changes are seen in $\alpha_{ox}$, demonstrating that when the spectrum appears
less complex, X-ray emission is stronger and $\alpha_{ox}$ is more typical.

The transiting of objects from the C- to the S-sample with increasing
flux is predicted by both partial-covering and reflection models.
In terms of partial-covering
the increased spectral complexity during the low-flux states 
occurs because
the intrinsic spectrum (presumably a blackbody plus power law)
is highly absorbed, imposing edges and curvature in the detected spectrum.
For reflection, the low-flux spectrum will likely be reflection dominated and
the spectral curvature and drops will be associated with various reflection components
(predominately the blurred iron line and soft excess; e.g. Ross \& Fabian 2005).

\subsubsection{Absorption}
The X-ray weakness is, of course, in agreement with absorption models such
as partial-covering where the X-ray emitting region is more absorbed than
the UV region.  This could occur, for example, if the absorber is located somewhere between
the X-ray and UV emitting regions of the accretion disc.  Consequently,
UV fluctuations could arise from e.g. changes in the accretion rate, while the
X-rays variations will appear larger because of the addition affect of the changing
absorption.  Therefore the extreme changes in the X-ray flux due to absorption
can drive the fluctuations in $\alpha_{ox}$
when compared to the relatively unabsorbed and less variable UV flux.
We also note that in the multi-epoch observations of \1h\ and NGC~4051 the UV
flux diminished when the X-ray flux was higher.

A prediction of the partial-covering model (e.g. Tanaka \et 2004) is that
changes in the covering fraction or column density of the absorber(s) will
account for, at least, the long-term spectral variations; but that the
unabsorbed, intrinsic flux will remain relatively constant over time.
We can test this prediction by reexamining the two \xmm\ observations of
\1h\ (Gallo \et 2004a).  During both observations, the absorbed component of the
intrinsic spectrum was
seen through a column density of $\gs10^{23}\pscm$; however during the first,
low-flux state observation the covering fraction\footnote{Note that the covering
fraction and column density are degenerate in partial-covering models and
cannot be distinguished.} of the absorber was much greater.
The intrinsic, $2\keV$ flux density during the first and second observation
of \1h\ was $13.13$ and $8.87\times10^{-30}\ergpscmpspHz$, respectively.
This yields $\alpha_{ox}=-1.17$ at $both$ epochs, demonstrating that a common
intrinsic flux modified by partial obscuration is possible in accounting for
the differences in brightness over time.

\subsubsection{Reflection}
Steep and fluctuating $\alpha_{ox}$ is not inconsistent with models, such as
light bending (e.g. Miniutti \& Fabian 2004), which incorporate general
relativistic effects close to the black hole to account for some of the
X-ray variability.  In these models the low-flux, reflection dominated
spectrum results when most of the emission from the primary, power law
component is bent back toward the black hole and never reaches the observer.
In comparison with the relatively constant UV emission (most likely produced
by the colder accretion disc) $\alpha_{ox}$ will be smaller (X-ray weaker).  As
the emission from the power law continuum becomes important, the X-ray source
becomes brighter and $\alpha_{ox}$ begins to flatten.

\subsection{Optical \feii\ emission}
Not all of the NLS1 in the C-sample possess strong optical \feii\ emission
(specifically \feii/\hbeta), but it does appear that the most
extreme \feii\ emitters (e.g. \feii/\hbeta$\gs1.7$; \iras13224, \1h, PHL~1092)
are spectrally complex.

Firm conclusions cannot be drawn due to the small number statistics, but
this raises the possible conjecture that not all object which display
complex X-ray spectra possess strong iron emission, but possibly all objects with
extreme \feii\ emission do possess the mechanism to create high-energy
complexities.
A simple test would be to make X-ray observations of a few
extreme \feii\ emitters, say with \feii/\hbeta$\gs1.7$, as suggested by
Figure~\ref{fig:feii}, and determine if their high-energy spectra are
complex.

However, the fact that a source can transit from one sample to
another depending on its X-ray flux state (Section~3) raises further issues and
limits the conclusions we can make regarding \feii\ emission.
The primary issue is whether the \feii\ emission is variable and
if it depends on the X-ray continuum flux.  There are several
studies which show evidence for (e.g. Kollatschny \& Fricke 1985;
Kollatschny \et 2000; Vestergaard \& Peterson 2005; Wang, Wei \& He 2005)
and against (Goad \et 1999; Kollatschny \& Welsh 2001) significant
\feii\ variability.
Since the \feii\ measurements presented here are not contemporaneous
with the X-ray measurements, we can only speculate on a possible
connection between \feii\ strength and X-ray spectral complexity.

An alternative possibility is that the strongest \feii\ emitters simply
show the greatest variability.  If so, this makes it more likely to
catch the strongest \feii\ emitters in an extreme (low) flux state,
as opposed to an average NLS1.  
\iras13224, \1h\ and \phl\ are certainly know for there strong \feii\
emission and significant variability.
From the current
analysis it is difficult to distinguish the two theory.

\section{Summary}
\label{sect:conc}

We have investigated the nature of NLS1 which exhibit complexity in their
high-energy ($2.5-10\keV$) spectrum.  We identified seven out of twenty-eight
NLS1 whose high-energy spectrum show significant deviations from
a simple power law.  We compared multiwavelength properties of this complex
sample of seven NLS1 with the larger sample to investigate whether we could
identify any underlying physical processes that could be responsible for
the spectral complexity.  The main results follow.

\begin{itemize}
\item
Considering long-term X-ray variability of the sources, the complex sample 
of NLS1 (C-sample) seems representative of objects in X-ray low-flux states, whereas the
NLS1 in the general sample (S-sample) appear to be in a typical flux state.

\item
In cases where multi-epoch measurements of $\alpha_{ox}$ were possible, it
appeared that $\alpha_{ox}$ approached more typical values as the complexity in
the high-energy spectrum diminished.
For \1h\ and NGC~4051 (both in the complex sample) the variations in $\alpha_{ox}$
were extreme and while predominately due to variability in the X-rays, 
the UV flux did also change.  Specifically, in these two objects,
the UV flux actually diminished when the X-ray flux was high and the spectral complexity
less prominent.  This is not inconsistent with either partial-covering or
reflection scenarios.

\item
On average, the C-sample showed stronger \feii\ emission (\feii/\hbeta)
than the general sample.  Specifically, the three most extreme \feii\ emitters
(with \feii/\hbeta$>1.8$) of the twenty-eight NLS1, were all part of the C-sample
of only seven.  
This raises an intriguing possibility that we may be
able to identify complicated X-ray NLS1 based on the strength of the optical
\feii\ emission.
It is not certain from this current analysis if the possible connection
between X-ray spectral complexity and \feii\ strength is due to high iron
abundances or because strong \feii\ emitters may exhibit greater X-ray
variations and consequently are more likely to be caught in an extreme (low)
flux state.

\end{itemize}

Both absorption and reflection models predict the observed behaviour
in $\alpha_{ox}$.  It is interesting to note that in the case of \1h\ a
common, intrinsic $\alpha_{ox}$ is predicted by the partial-covering model
and is consistent with the \xmm\ observations.  

Partial-covering models make no strong assumption on the nature of the
intrinsic spectrum, only that it is partly obscured.  In this sense reflection
models, which can describe the broad-band X-ray properties are attractive.
While it seems obvious that reflection close to the black hole should occur
it also seems premature to abandon concepts which employ absorption in the black
hole environment to explain some of the X-ray behaviour.


\section*{Acknowledgements}

Many thanks to Todd Boroson and Dirk Grupe for providing the black
hole data.  I am very gratefully to Thomas Boller,
Niel Brandt, Andy Fabian, G\"unther
Hasinger and Yasuo Tanaka for comments on the manuscript and many helpful 
discussions over the years.
Much appreciation to the referee for a constructive report which has
resulted in an improved paper.
LCG acknowledges funding from the Japan Society for the Promotion of
Science through a JSPS Postdoctoral Fellowship.
Based on observations obtained with \xmm, an ESA science mission with
instruments and contributions directly funded by ESA Member States and
the USA (NASA). 
This research has made use of the Tartarus (Version 3.2) database, 
created by Paul O'Neill and Kirpal Nandra at Imperial College London, 
and Jane Turner at NASA/GSFC. Tartarus is supported by funding from 
PPARC, and NASA grants NAG5-7385 and NAG5-7067.

\bsp
\label{lastpage}
\end{document}